\RequirePackage{fix-cm}
\documentclass[twocolumn]{svjour3}          
\smartqed  
\usepackage{graphicx}
\usepackage{mathptmx} 
\usepackage{bm}
\usepackage{amsmath}
\usepackage{amssymb}
\usepackage{amsfonts}
\usepackage{textcomp}
\usepackage {ulem}
\usepackage{natbib}
\usepackage[usenames]{color}

\definecolor{green4}{RGB}{34, 177, 76}
\definecolor{green1}{RGB}{0, 204, 102}
\definecolor{green2}{RGB}{153, 255, 0}
\definecolor{orange1}{RGB}{255, 102, 0}
\definecolor{purple}{RGB}{153, 0, 255}
\definecolor{blue2}{RGB}{0, 102, 255}
\definecolor{green3}{RGB}{0, 204, 0}
\definecolor{orange2}{RGB}{255, 153, 0}

\journalname{Microfluidics and nanofluidics}
\begin{document}

\title{Secondary flows of viscoelastic fluids in serpentine microchannels
}


\author{Lucie Duclou\'e \and Laura Casanellas \and Simon J. Haward \and Robert J. Poole \and Manuel A. Alves \and Sandra Lerouge \and Amy Q. Shen \and Anke Lindner
}


\institute{L. Duclou\'e \and L. Casanellas \and A. Lindner  \at
              Laboratoire de Physique et M\'ecanique des 
              Milieux H\'et\'erog\`enes, UMR 7636, CNRS, ESPCI Paris, PSL Research University,
              Universit\'e Paris Diderot, Sorbonne Universit\'e,  
              Paris, 75005, France \\
              \email{anke.lindner@espci.fr}             \\
             \emph{Present address:} of L. Casanellas, Laboratoire Charles Coulomb UMR 5221 CNRS-UM, Universit\'e
de Montpellier, Place Eug\`ene Bataillon, 34095 Montpellier CEDEX 5, France
           \and
           S.J. Haward \and A.Q. Shen \at
              Okinawa Institute of Science and Technology Graduate University, Onna, Okinawa 904-0495, Japan
           \and
           R.J. Poole \at
          School of Engineering, University of Liverpool, Brownlow Street, Liverpool L69 3GH, United Kingdom
          \and
          M.A. Alves \at
          Departamento de Engenharia Qu\'{i}mica, Centro de Estudos de Fen\'{o}menos de Transporte, Faculdade de Engenharia da Universidade do Porto, Rua Doutor Roberto Frias, 4200-465 Porto, Portugal
          \and
          S. Lerouge \at
          Laboratoire Mati\`ere et Syst\`emes Complexes, CNRS UMR 75057-Universit\'{e} Paris Diderot, 10 rue Alice Domond et L\'{e}onie Duquet, 75205 Paris CEDEX, France
}

\date{Received: date / Accepted: date}

\maketitle

\begin{abstract}


Secondary flows are ubiquitous in channel flows, where small velocity components perpendicular to the main velocity appear due to the complexity of the channel geometry and/or that of the flow itself such as from inertial or non-Newtonian effects, etc. We investigate here the inertialess secondary flow of viscoelastic fluids in curved microchannels of rectangular cross-section and constant but alternating curvature: the so-called ``serpentine channel" geometry. Numerical calculations (Poole et al, 2013) have shown that in this geometry, in the absence of elastic instabilities, a steady secondary flow develops that takes the shape of two counter-rotating vortices in the plane of the channel cross-section. We present the first experimental visualization evidence and characterization of these steady secondary flows, using a complementarity of \textmu PIV in the plane of the channel, and confocal visualisation of dye-stream transport in the cross-sectional plane. We show that the measured streamlines and the relative velocity magnitude of the secondary flows are in qualitative agreement with the numerical results. In addition to our techniques being broadly applicable to the characterisation of three-dimensional flow structures in microchannels, our results are important for understanding the onset of instability in serpentine viscoelastic flows.

\end{abstract}

\section{Introduction}
\label{intro}

Three-dimensional velocity fields are widespread in channel and pipe flows, where the geometry of the duct can combine with the properties of the base primary flow (i.e. the flow in the streamwise direction) to trigger a weak current with velocity components perpendicular to the streamwise direction. Therefore, the ability to measure, and understand, the velocity field in all three directions of such flows is of general importance. In microfluidic flows, however, the flowfield in the streamwise direction is often the only component characterised, because of optical access limitations and due to the fact that the absolute value of the other velocity components are typically very small~\citep{Tabeling2005}. Despite their small magnitude, such secondary flows are often ultimately responsible for enhanced mixing (above that due to diffusion alone) of mass and heat which is a frequent aim of various microfluidic devices~\citep{lee2011, Mitchell2001, Stroock2002, Amini2013, Hardt2005, Kockmann2003}. Secondary flows also have important implications in particle focussing \citep{Delgiudice2015, DiCarlo2007} where they may either be exploited or act as a hindrance.

Secondary flows in microfluidic systems may be driven by the complexity of the channel geometry only: for the creeping flow of a Newtonian fluid, \cite{Lauga2004} have shown that a secondary flow must develop if the channel has both varying cross-section and streamwise curvature. Changes in the streamwise curvature of channels with constant cross-section have also been shown to give rise to a secondary flow around the bend~\citep{guglielmini2011, Sznitman2012}. More complex geometries have been designed to obtain chaotic micromixers, in which secondary flows are triggered and expose volumes of fluid to a repeated sequence of rotational and extensional local flows~\citep{Ottino1989, Stroock2002, Amini2013}. 

Complexity in the equations of fluid motion is another driving mechanism for secondary flows: although usually not dominant at the microscale, inertia can play a role in microfluidic systems~\citep{di2009, Amini2014}. Combined with the flow geometry, it drives secondary flows such as the well-known ``Dean'' vortices~\citep{Dean1927, Dean1928} observed in curved channels and pipes, or the steady vortical structure of the ``engulfment'' regime in T-junction mixers \citep{Kockmann2003, fani2013}. In the absence of inertia, viscoelasticity is another source of fluid dynamic complexity: viscoelastic analogues of the Dean vortices are formed, in the creeping-flow regime, by the coupling of the first normal-stress difference with streamline curvature~\citep{Robertson1996, fan2001, Poole2013, Bohr}. Note that second-normal stress differences in viscoelastic fluids may also drive an inertialess secondary motion in ducts of non-axisymmetric cross-section, but this flow is typically much weaker~\citep{Gervang1991, Debbaut1997, Xue1995}. We emphasise that in listing those potential sources of secondary flow we are not attempting to be exhaustive, but simply to illustrate that they may occur under many different scenarios.

We focus here on the viscoelastic secondary flow driven by streamline curvature. This steady secondary flow is always present in the steady flow of viscoelastic fluids in curved geometries, and pertains at all flow rates until a critical flow rate is reached at which the flow becomes time-dependent due to a well-known purely elastic instability~\citep{Groisman2000, Arratia2006, Afik2017, Soulies2017}. Characterising this secondary flow is thus essential to the knowledge of the three-dimensional base flow from which the elastic instability develops: its structure may interact with the onset of the instability, as hypothesised to explain the partially unaccounted for stabilisation of shear-thinning viscoelastic flow in curved microchannels~\citep{Casanellas2016}. It is also important for mixing and particle focusing applications that rely on viscoelastic fluids~\citep{Groisman2000, Delgiudice2015}.

\begin{figure}[h!]
\begin{center}
  \includegraphics[width=0.95\linewidth]{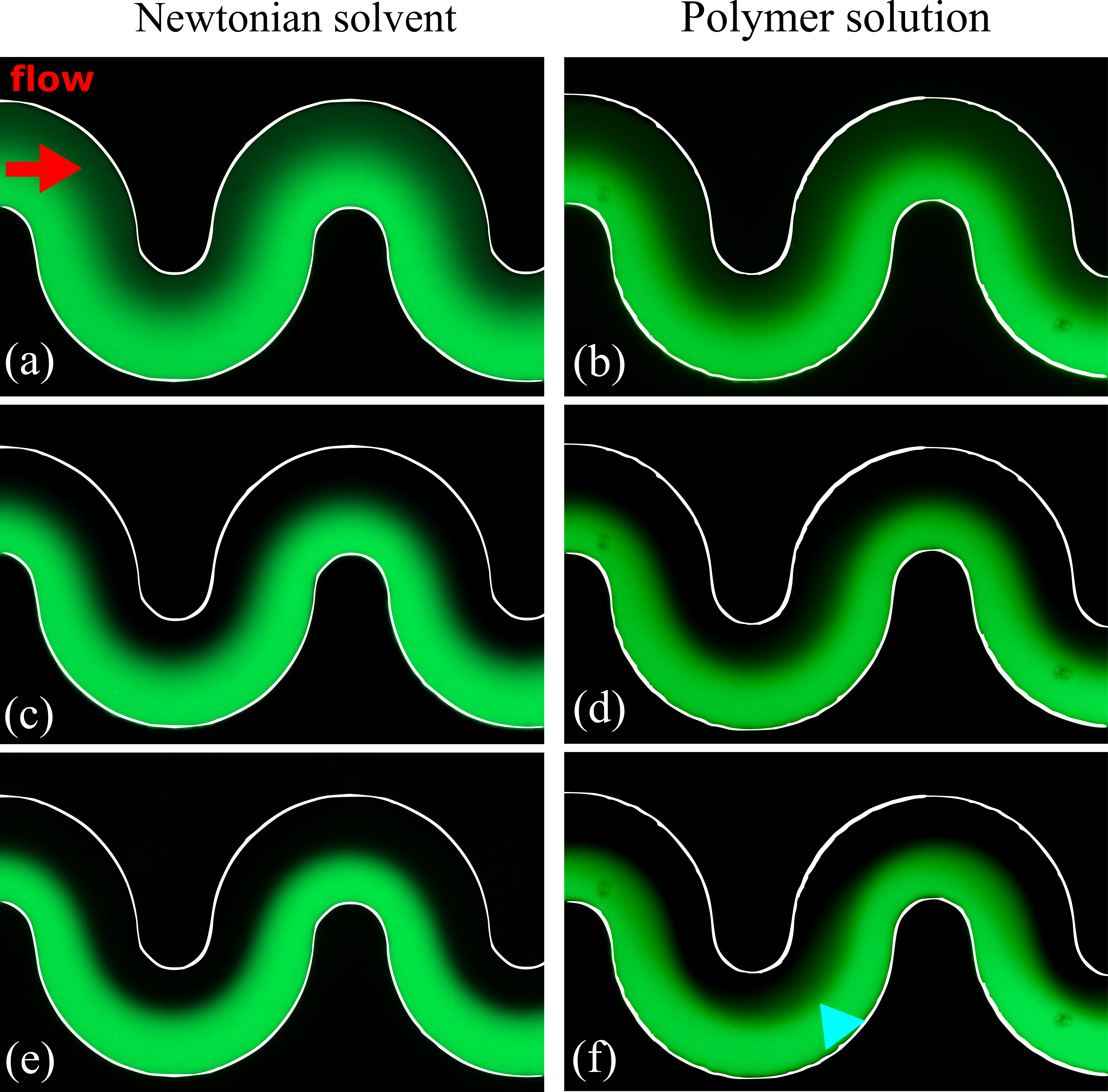}
\caption{Visualisation of mixing in a serpentine microchannel (the channel edges are highlighted in white): two streams of fluid are co-injected in a Y-junction, one of them being fluorescently labelled. Data for a viscoelastic polymer solution are displayed on the right-hand side, while data for the Newtonian solvent (a mixture of water and glycerol at 75~wt.\% -- 25 wt.\%)\protect\footnotemark ~are shown on the left-hand side. The flow rate increases from 2~\textmu l/min (top row) to 6~\textmu l/min (middle row) and 12~\textmu l/min (bottom row).  At low flow rates ((\textit{a}), (\textit{b})) the interface between the two streams is broadened in both cases by the strong diffusion of the dye. At larger flow rates ((\textit{c}) and (\textit{d})), the interface sharpens all along the channel due to the decreasing residence time. Further increase of the flow rate ((\textit{e}) and (\textit{f})) leads to further sharpening of the interface for the Newtonian flow (\textit{e}), but an additional spatially varying ``blur" develops in the viscoelastic flow ((\textit{f}), blue triangular arrow).}
\label{fig:1}
\end{center}
\end{figure}

Evidence for such secondary flows is readily observable in simple visualisation experiments. By way of example, in Fig.\ref{fig:1} we show a classical experiment for the visualisation of mixing efficiency in a serpentine microchannel~\citep{Groisman2000}: two streams of the same fluid, one of them dyed with fluorescein, are co-injected into the serpentine micromixer. When a Newtonian fluid is injected, mixing is achieved by diffusion alone, which broadens the interface. With increasing flow rate, the residence time decreases, and so does the width of the interface. When a viscoelastic fluid is used, the evolution of the width of the interface with increasing flow rate is very different. At small flow rates a very broad interface is again observed, which initially sharpens when the flow rate is increased. When the flow rate is further increased however, the interface locally widens again. This effect cannot be attributed to diffusion, which becomes less important with increasing flow rate (thus decreasing residence times). In addition, an asymmetry can be observed with the interface being significantly wider towards the end of each loop and a sharpening of the interface at the beginning of each new loop. This observation can only be explained with an underlying three dimensional flow structure that reverses direction in between consecutive loops. Previous numerical simulations \citep{Poole2013} have shown the occurrence of a steady secondary flow in this serpentine channel geometry for dilute viscoelastic liquids. One of the aims of this work is to demonstrate and quantify the occurrence of this secondary flow experimentally by direct measurement. More generally, we show how different experimental methods can be used to determine quantitative information of generic secondary flows in micro-devices.

\footnotetext{The small fluorescein molecule diffuses almost freely in the polymer network and thus probes a local viscosity that is lower than the shear viscosity of the polymer solution as measured with a rheometer. The same behaviour has been quantified in solutions of (hydroxypropyl)cellulose~\citep{Mustafa1993}, dextran~\citep{Furukawa1991} and polyethylene glycol~\citep{Holyst2009}. For this reason, we use the solvent of the polymer solution as a Newtonian reference fluid. The slightly lower diffusion in the polymeric solution shows that the contribution from the polymer to the local viscosity, albeit small, is not entirely negligible.}

Being typically very weak (on the order of a few percent of the bulk primary velocity), secondary flows are hard to resolve even in macro-sized classical fluid mechanics experiments~\citep{Gervang1991}. Thus it is not surprising that such flows have been little characterised at the microscale. A number of recent experimental approaches may alleviate this issue, in particular the holographic microparticle tracking velocimetry (\textmu PTV) technique~\citep{Salipante2017}, confocal microparticle image velocimetry (confocal \textmu PIV)~\citep{li2016} or using standard particle image velocimetry in conjunction with a channel design and material that allow for microscope observation in several perpendicular planes~\citep{Burshtein2017}.

Here, we will characterise experimentally the three-dim\-ensional structure of the flow with supporting numerical simulations that match the geometrical conditions. Our aim is to use a complementarity of \textmu PIV, confocal microscopy and insight gleaned from simulation to quantify the secondary flow and confirm its vortical structure and sense of rotation. Our techniques are very generic and thus broadly applicable to the characterisation of three-dimensional flow structures in microchannels.\\

\section{Experimental and numerical methods}
\label{sec:mats-meths}
\subsection{Working fluids and rheological characterisation}
\label{sec:exp}

Model viscoelastic fluids were prepared by dissolving poly\-ethylene oxide (PEO, from Sigma Aldrich) with a molecular weight of $M_W=4 \times 10^6$ g/mol in a water/glycerol (75\% - 25\% in weight) solution. The PEO was supplied from the same batch as used in~\cite{Casanellas2016}. The solvent viscosity at $T=21^{\circ}$C is $\eta_s= 2.1$ mPa$\cdot$s (data not shown). The polymer concentration was fixed to $c=500$ ppm (w/w). The total viscosity of the resulting solution at $T=21^{\circ}$C is $\eta= 3.8$ mPa$\cdot$s giving a solvent-to-total viscosity ratio $\beta$ = 0.55. The overlap concentration for this polymer in water is $c^*\simeq 550$ ppm \citep{Casanellas2016}. Although this solution is close to the semi-dilute limit, we confirmed that shear-thinning effects, of both the shear viscosity and the first normal-stress difference, are essentially negligible (see e.g. \cite{Casanellas2016}).


\subsection{Microfluidic geometry}
\label{sec:expgeom}
\begin{figure}[h]
  \includegraphics[width=\linewidth]{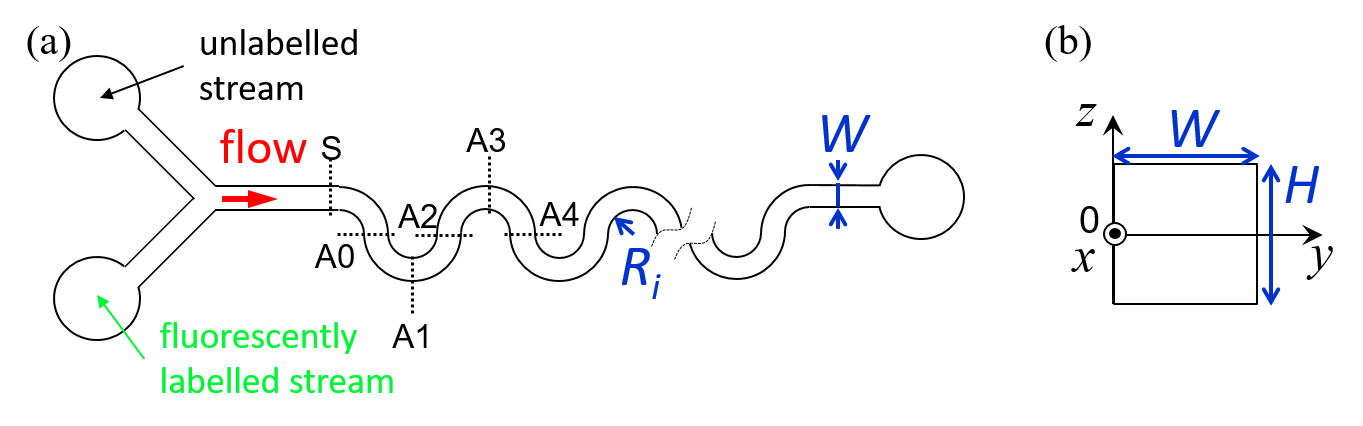}
\caption{Schematic of the microfluidic geometry used: \textit{(a)} top view and \textit{(b)} cross-sectional view displaying the choice of axes. $x$ is the primary flow velocity direction, $y$ is the wall-normal direction (where the origin is taken at the inner edge of each loop), and $z$ is the spanwise (vertical) direction. Therefore, the $x, y, z$ coordinate system we consider is not fixed in space but advected with the flow. The location of the dyed stream used for confocal visualisation is also indicated.}
\label{fig:2}
\end{figure}

We tested the polymer solution in serpentine microchannels consisting of nine half loops. A sketch of the channel is shown in Fig.\ref{fig:2}. We note that, in this channel geometry, the absolute value of the curvature is constant along the channel but the sign of the curvature changes from positive to negative between consecutive half-loops. This change of sign is not required for the development of the viscoelastic secondary flow, but is a feature of our two-dimensional geometry which conveniently allows for the study of several consecutive loops at a constant radius of curvature. Numerical simulations of the creeping flow of Newtonian fluids in bent microchannels show that this change in curvature is expected to trigger a local secondary flow where subsequent half-loops reconnect, but this flow quickly decays in the regions of constant curvature~\citep{guglielmini2011}, where our velocity measurements were made. Most of our results were obtained on a channel of nearly square cross-section, with a width $W=110 \pm 3$~\textmu m, height $H=99 \pm 1$~\textmu m and an inner radius of curvature (measured at the inner wall of the channel) $R_i = 40 \pm 1$~\textmu m. Additional channels of comparable cross-sectional dimensions but larger radii of curvature were used for comparison. 

The microchannels were fabricated in polydimethylsiloxane (PDMS), using standard soft-lithography microfabrication methods~\citep{Tabeling2005}, and mounted on a glass coverslip. The fluid was injected into the channel via two inlets using two glass syringes (Hamilton, 500~\textmu l each) that were connected to a high-precision syringe pump (Nemesys, from Cetoni GmbH). The experimental protocol consisted of stepped ramps of increasing flow rate from 2~\textmu l/min up to a maximum of 20~\textmu l/min, with a flow rate step of 2~\textmu l/min. The resolution of the applied flow rate was controlled at a precision of $\pm$ 0.2 \textmu l/min, as confirmed independently using a flow sensor (Flow unit S, from Fluigent, at low flow rates and SLI-0430 Liquid flow meter, from Sensirion for $Q \geqslant 6$~\textmu l/min). The step duration was set to 120~s, and the measurements performed over the last 60~s, which we confirm was long enough to ensure flow steadiness and the decay of any initial transient regime. Experiments were continued until the onset of the purely-elastic instability where the flow became time-dependent. For the $R_i = 40$~\textmu m channel this occurred at 14~\textmu l/min which we define as $Q_{insta}$. 

At the onset of flow instability the Reynolds number ($Re= \rho UW/\eta$, where $\rho$ is the fluid density and $U$ the mean velocity) is small in all experiments, being at most 0.6. Therefore, in our microfluidic flow experiments inertial effects can be disregarded.

\subsection{Micro-particle image velocimetry}

Quantitative two-dimensional measurements of the flow field were made in the $xy$-centreplane ($z=0$ plane) of the serpentine device (Fig.\ref{fig:2}) using a micro-particle image velocimetry (\textmu PIV) system (TSI Inc., MN)~\citep{Meinhart2000, Wereley2005}. For this purpose, no fluorescent dye was used but the test fluid was seeded with 0.02~wt\% fluorescent particles (Fluoro-Max, red fluorescent microspheres, Thermo Scientific) of diameter $d_p = 0.52$~\textmu m with peak excitation and emission wavelengths of 542~nm and 612~nm, respectively. The microfluidic device was mounted on the stage of an inverted microscope (Nikon Eclipse T\textit{i}), equipped with a 20$\times$ magnification lens (Nikon, $\textnormal{NA}=0.45$). With this combination of particle size and objective lens, the measurement depth over which particles contribute to the determination of the velocity field was $\delta z_m \approx 13$~\textmu m~\citep{Meinhart2000}, which is approximately 13\% of the channel depth.

The \textmu PIV system was equipped with a $1280 \times 800$ pixel high speed CMOS camera (Phantom MIRO, Vision Research), which operated in frame-straddling mode and was synchronized with a dual-pulsed Nd:YLF laser light source with a wavelength of  527~nm (Terra PIV, Continuum Inc., CA). The laser illuminated the fluid with pulses of duration $\delta t <10$~ns, thus exciting the fluorescent particles, which emitted at a longer wavelength. Reflected laser light was filtered out by a G-2A epifluorescent filter so that only the light emitted by the fluorescent particles was detected by the CMOS imaging sensor array. Images were captured in pairs (one image for each laser pulse), where the time between pulses $\Delta t$ was set such that the average particle displacement between the two images in each pair was around 4 pixels. Insight 4G software (TSI Inc.) was used to cross-correlate image pairs using a standard \textmu PIV algorithm and recursive Nyquist criterion. The final interrogation area of 16 $\times$ 16 pixels provided velocity vector spaced on a square grid of 6.4 $\times$ 6.4 \textmu m in $x$ and $y$. The velocity vector fields were ensemble-averaged over 50 image pairs. 

\subsection{Confocal microscopy}
Vertical images of the cross-section of the channel (in $yz$ planes) were obtained by confocal microscopy imaging using dyed stream visualisation (as illustrated in the $xy$ plane in Fig.\ref{fig:1}). $z$-stacks of two-dimensional images of size 1024$\times$1024 pixels in $xy$ planes were acquired at a rate of $6$~fps using a laser line-scanning confocal fluorescence microscope (LSM 5 Live, Zeiss), with a 40$\times$ water immersion objective lens (1.20 NA). The voxel size was 0.16 $\times$ 0.16 $\times$ 0.45~\textmu m in the $x - y - z$ direction.

\subsection{Viscoelastic constitutive equation, numerical method and structure of predicted secondary flow}
\label{sec:num}

The three-dimensional numerical simulations assume isothermal flow of an incompressible viscoelastic fluid described by the upper-convected Maxwell (UCM) model~\citep{Oldroyd1950} in a channel of matched dimensions to those used in the experiments. The equations that need to be solved are those of mass conservation,

\begin{equation}
\label{eqn1}
\nabla \cdot \textbf{u} = 0,
\end{equation} 

and the momentum balance,

\begin{equation}
\label{eqn2}
- \nabla p + \nabla \cdot \tau = \mathbf{0},
\end{equation}

assuming creeping-flow conditions (i.e. the inertial terms are exactly zero), where $\mathrm{\mathbf{u}}$ is the velocity vector with Cartesian components ($u_x$, $u_y$, $u_z$), and $p$ is the pressure. For the UCM model the evolution equation for the polymeric extra-stress tensor, $\tau$, is

\begin{equation}
\label{eqn3}
\boldsymbol \tau + \lambda \boldsymbol \tau_{(1)}=\eta \dot{\boldsymbol \gamma},
\end{equation} 

where $\boldsymbol\tau_{(1)}$ represents the upper-convected derivative of $\boldsymbol \tau$ and $\eta$ the constant polymeric contribution to the viscosity of the fluid, respectively. 

Although the UCM model exhibits an unbounded steady-state extensional viscosity above a critical strain rate (1/2$\lambda$), in shear-dominated serpentine channel geometries such model deficiencies are unimportant and it is arguably the simplest differential constitutive equation which can capture many aspects of highly-elastic flows. Many more complex models (e.g. the FENE-P, Giesekus and Phan -- Thien -- Tanner models, see e.g.~\cite{Bird1987}), simplify to the UCM model in certain parameter limits and thus its generality makes it an ideal candidate for fundamental studies of viscoelastic fluid flow behaviour. The governing equations were solved using a finite-volume numerical method, based on the logarithm transformation of the conformation tensor. Additional details about the numerical method can be found in \cite{Afonso2009, Afonso2011} and in other previous studies (e.g. \cite{Alves2003a, Alves2003b}). For low Deborah numbers $De = \lambda U/R_i$ (with $\lambda$ the relaxation time of the fluid, $U$ the mean velocity and $R_i$ the inner radius of curvature), the numerical solution converges to a steady solution, which was assumed to occur when the $L^1$ norm of the residuals of all variables reached a tolerance of $10^{-6}$. Beyond a critical Deborah number, a time-dependent purely-elastic instability occurs. The numerical results in the current paper are restricted to Deborah numbers below the occurrence of this purely-elastic instability, thus the flow remains steady in all simulations.

\begin{figure}[h]
  \includegraphics[width=1\linewidth]{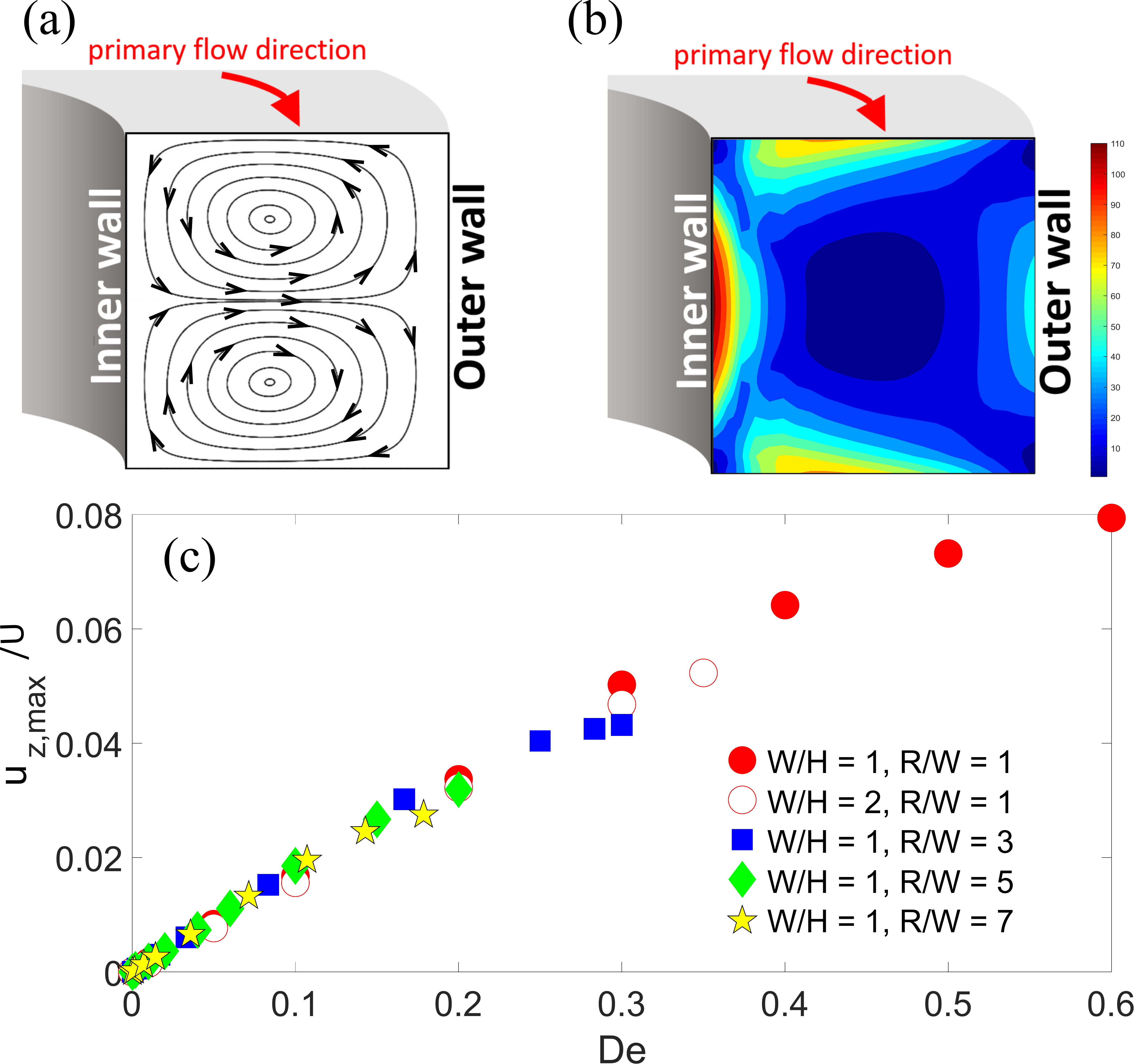}
\caption{\textit{(a)} Numerically predicted structure of the secondary flow: two counter-rotating vortices develop in the cross-sectional plane of the channel. \textit{(b)} Contour plot of the first normal stress difference ($N_1/(\eta U/W)$), in the cross-section of the channel ($De = 0.5$, $W/H=W/R=1$): the strong positive and asymmetric normal stress difference at the top and bottom wall drives the secondary flow. \textit{(c)} Numerically calculated scaling of the strength of the secondary flow: the maximum spanwise velocity scales linearly with $De$ over a wide range of aspect ratio and radius of curvature parameters (adapted from~\cite{Poole2013}).}
\label{fig:3}
\end{figure}

The structure and strength of viscoelastic secondary flows in a serpentine geometry have been investigated in detail by~\cite{Poole2013}. The main results of this study are recalled below, and lay the foundations for the simulations that we carry out here, which are focused on the geometry of the experimental system we used. The projected streamlines in the $yz$ plane of the computed secondary flow are shown in Fig.\ref{fig:3}(a). The flow takes the shape of a pair of counter-rotating vortices in the cross-sectional plane of the channel. It is driven by the hoop stress, which drives the fluid towards the inner side of the bend close to the top and bottom walls (where the shear rate and thus the first normal stress difference are larger, as illustrated in Fig.\ref{fig:3}(b)). The fluid is then carried back towards the outer edge of the bend at the centreplane ($z = 0$). Although the driving mechanism is different, the resulting qualitative features of this elasticity-driven secondary flow are thus similar to the inertia-driven Dean vortices~\citep{Dean1928}. The strength of the viscoelastic secondary flow increases with the elastic contribution to the flow (increasing $De$) and the curvature of the channel. \cite{Poole2013} have shown that far from the onset of the purely elastic instabilities, the magnitude of the secondary flow, as quantified by the maximum spanwise velocity $u_{z, max}$, scales linearly with $De$ (see Fig.\ref{fig:3}(c)). The structure of the flow has been shown to remain identical for aspect ratios $W/H$ varying from 1 up to 4; accordingly, the scaling for $u_{z, max}$ with $De$ is not modified by the aspect ratio of the channel. The scaling of the secondary flow strength with the solvent viscosity contribution has also been assessed~\citep{Poole2013}, and can be expressed as an effective Deborah number $De_{eff} = (1- \beta) De$ where $\beta$ is the ratio of the solvent viscosity to the total (solvent + polymeric) viscosity. In the UCM model, $De_{eff}=De$. 

The precise experimental determination of the relaxation time of the fluid is difficult for dilute polymer solutions so that the determination of $De$ for our experimental data is challenging. However, we can use the onset of the purely elastic instability as a reference point to match the Deborah numbers in our numerical and experimental data. For a given serpentine geometry, the flow becomes unstable beyond a critical flow speed, usually expressed in terms of a critical Weissenberg number to quantify the importance of the elastic contribution to the flow: $Wi_{insta} (R_i) = \lambda U_{insta}/W$ (with $Wi = De \times R_i/W$ the Weissenberg number)~\citep{Zilz2012}. For a given channel geometry $Wi/Wi_{insta} = De/De_{insta} = Q/Q_{insta}$. Therefore, to enable quantitative comparison between experimental and numerical results, all the results for the channel geometry described above will be presented in terms of the reduced quantity $De^* = De/De_{insta} = Q/Q_{insta}$, with $De_{insta} \approx 1.24$ from the numerical results for the experimental geometry we used ($R_i/W = 0.36$). $De^*$ is independent of $\beta$ and therefore, the solvent viscosity does not need to be taken into account in the present numerical simulations.

\section{Results}
\label{sec:results}
\subsection{Flow measurement in the plane of the channel using \textmu PIV}
\label{sec:PIV}

\begin{figure}[h]
\begin{center}
  \includegraphics[width=0.7\linewidth]{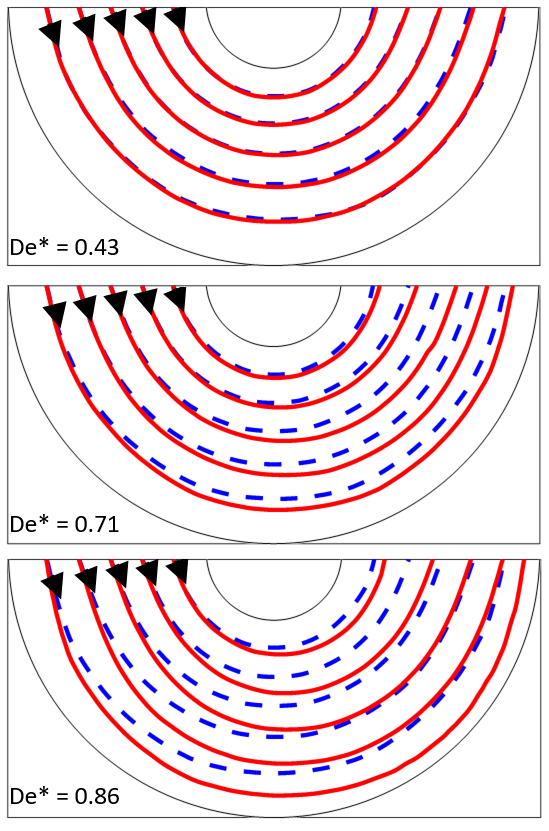}
\caption{Red lines: Experimental flow streamlines in the channel centreplane, for increasing values of flow elasticity. For comparison, the streamlines for a Newtonian solution of the same viscosity are shown with a blue dashed line: a clear deviation of the streamlines towards the outer edge of the bend occurs in viscoelastic flow.}
\label{fig:4}
\end{center}
\end{figure}

In this section we show that classical \textmu PIV measurements in the $xy$ centreplane provide significant evidence of the existence of a secondary flow. In this symmetry plane, $u_z = 0$ so that the velocity field is fully characterised by the 2D PIV. In Fig.\ref{fig:4} we show streamlines constructed from the measured velocity field along the first half loop (i.e. from A0 to A2 as shown in Fig.\ref{fig:2}). The blue dashed line highlights the Newtonian result which can be seen to travel in approximate concentric semicircles around the bend. The red lines indicate the streamlines of the polymeric solution, which, at low Deborah number, can be seen to match the Newtonian ones very closely. In contrast, with increasing $De^*$ (increasing flow rate) there is a marked deviation of the streamlines for the viscoelastic fluid \textit{away} from the inner bend towards the outer, in good agreement with the sense of the secondary flow predicted by the numerical simulations~(\cite{Poole2013}) and as discussed above and shown in Fig.\ref{fig:3}. These streamline patterns thus provide our first piece of qualitative evidence for the existence of an elastic secondary flow.

\begin{figure}[h!]
  \includegraphics[width=\linewidth]{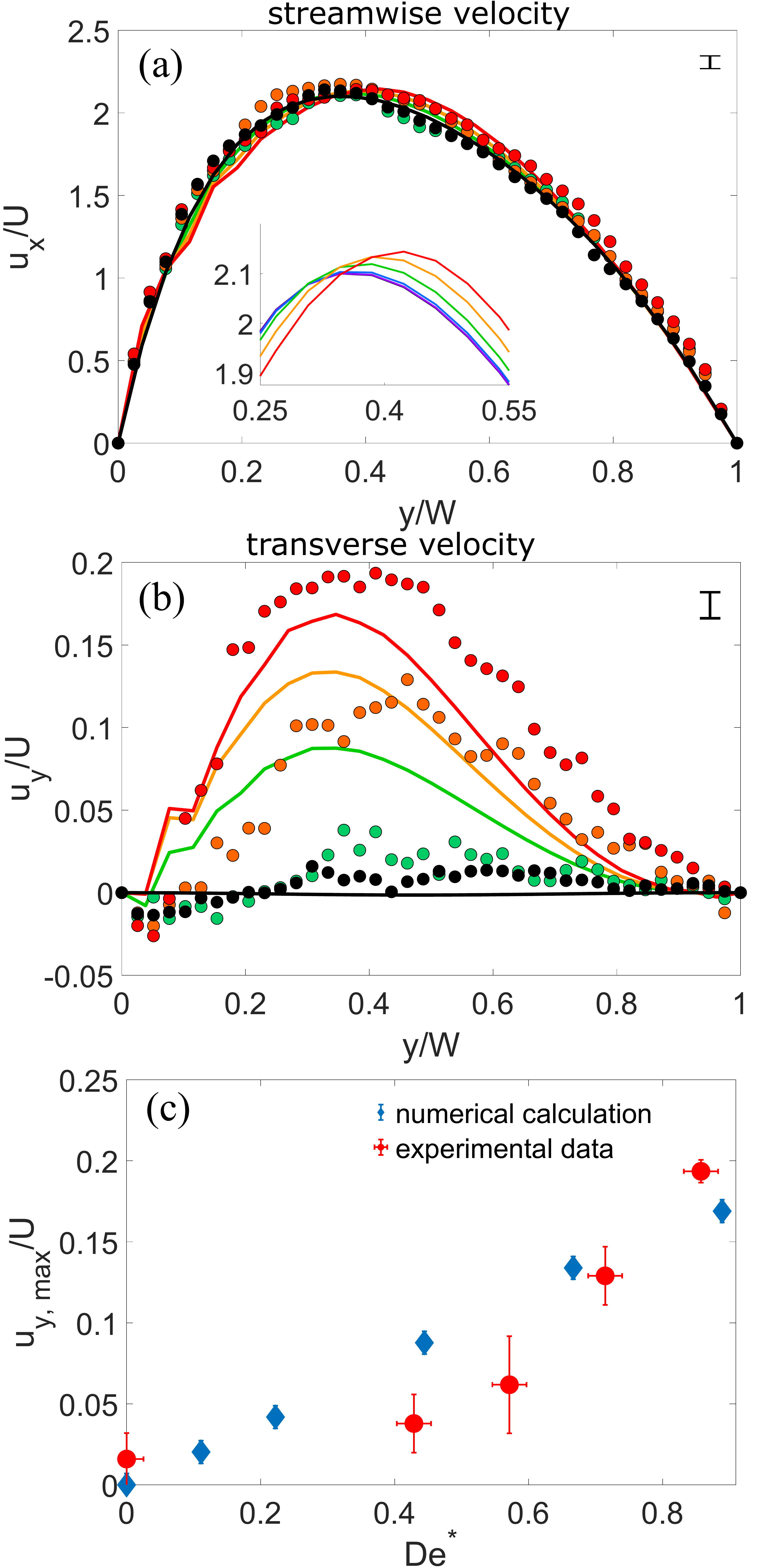}
\caption{Experimental (bullets) and numerical (full lines) velocity plots at location A1 (see Fig.\ref{fig:2}) for the primary (\textit{a}) and wall-normal velocity (\textit{b}), for increasing values of $De^*$ (\textcolor{black}{$\bullet$} Newtonian fluid, \textcolor{green1}{$\bullet$} 0.43, \textcolor{orange1}{$\bullet$} 0.71, \textcolor{red}{$\bullet$} 0.86 and \textcolor{black}{\textbf{-}} 0, \textcolor{purple}{\textbf{-}} 0.11, \textcolor{blue2}{\textbf{-}} 0.22, \textcolor{green3}{\textbf{-}} 0.44, \textcolor{orange2}{\textbf{-}} 0.67, \textcolor{red}{\textbf{-}} 0.89). For easier comparison, only select $De^*$ values are shown in \textit{(b)}. The zero $y$ location is taken at the inner edge of the bend, with the geometry being that described in section~\ref{sec:expgeom}: $R_i/W = 0.36$. The peak streamwise velocity shifts with $De$, as is more visible in the inset of \textit{(a)}. The scale bar in the top right corner indicates the magnitude of the error on the experimental data. (\textit{c}) Scaling for the maximum wall-normal velocity with $De^*$.}
\label{fig:5}
\end{figure}
To support these qualitative streamline observations in a more quantitative sense, in Fig.\ref{fig:5} we plot the velocity components around location A1. The primary and wall-normal components of both the experimentally determined (symbols) and the numerically computed (full lines) velocity fields have been averaged over an angular sector of $10^{\circ}$ upstream and downstream of A1. $y$ describes as usual the wall-normal direction, and $x$ the primary velocity direction: the primary velocity component is $u_x$ and the transverse (or wall normal) component is $u_y$. The data is plotted such that the zero $y$ location corresponds to the inner edge at location A1. In Fig.\ref{fig:5}(a) we plot the streamwise velocity component. Firstly we notice the good agreement between the experiments and the Newtonian simulation (black line and black symbols) where a slight asymmetry in the profile towards the inner wall is noticeable (as has been observed and discussed previously~\citep{Zilz2012}). Secondly, the effect of elasticity on this main velocity component is rather subtle but, as is most easily seen via inspection of the numerical profiles around the maximum of $u_x$ (see inset of Fig.\ref{fig:5}(a)), it is clear that elasticity acts to reduce this asymmetry by shifting the peak velocity back towards the centre of the channel. We then turn to the wall-normal velocity component, shown in Fig.\ref{fig:5}(b). For the Newtonian case this is essentially zero (within $\pm$ 1\% of the bulk velocity) - in agreement with theoretical predictions for an inertialess duct of constant cross-sectional area and constant curvature~\citep{Lauga2004}. However, with the polymer solution, increasingly large wall-normal velocities (i.e. from the inner wall towards the outer) can be discerned as the flow rate (or $De^*$) is increased. At the highest $De^*$ for both simulation and experiment these velocities reach $\approx$ 0.15$U$ at their peak. It can be observed that, much as is the case for the primary velocity component, these transverse velocity profiles are also asymmetric with a peak closer to the inner wall. We believe that this is a consequence of the shear rate being larger close to the inner wall. As the streamwise normal stress -- which, in combination with streamline curvature, is the driving force for the secondary flow -- increases with the shear rate, the higher shear rate at the inner wall leads to a concomitant asymmetry in the distribution of the secondary flow. Significant noise is visible on the experimental data, which is due to the difficulty of resolving accurately a velocity component much smaller than the average velocity. The systematic small discrepancies between the numerically computed profiles and the experimental data may be caused by the uncertainty on $Q_{insta}$ (known with a precision of $\pm 1$~\textmu l/min).

We quantify in Fig.\ref{fig:5}(c) the increase in the magnitude of the secondary flow with the Deborah number $De^*$. The maximum of the numerically computed transverse velocity $u_y$ scales linearly with $De^*$ over the range of parameters considered, as expected far from the instability onset~\citep{Poole2013}. The trend in the experimental data is more difficult to resolve because of the noise level, which is particularly high compared to the expected velocities at the lower flow rates investigated. Qualitative agreement is nonetheless observed, with comparable magnitude for the secondary flow in both cases.

\subsection{Cross-sectional visualisation of the flow using confocal microscopy}
\label{sec:confocal}

\begin{figure*}[h!]
\includegraphics[width=\textwidth=]{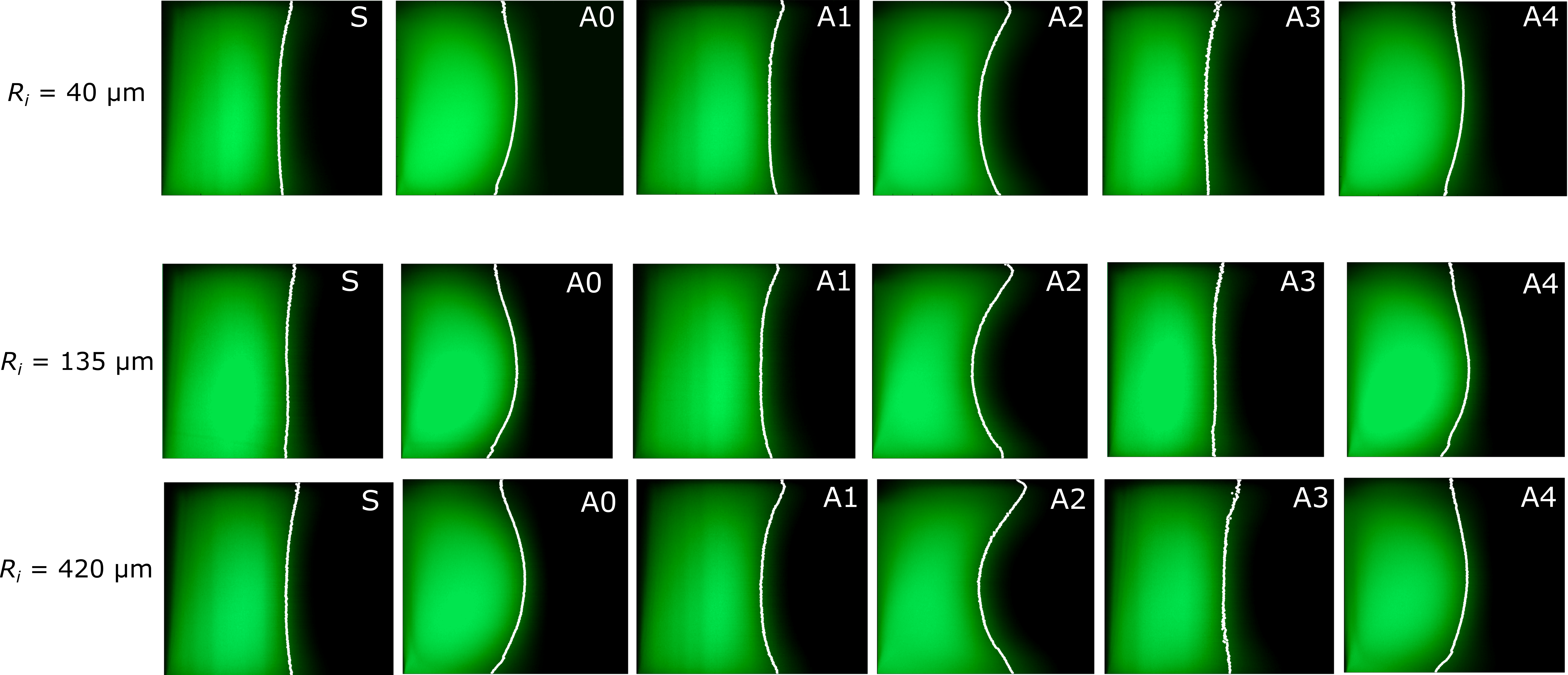}
\caption{Top row: Evolution of the cross-sectional view down the channel for the polymeric solution flown at $De^*=0.86$ in the channel used for the \textmu PIV experiments: the white line highlights the position of the interface, as obtained by adjusting the horizontal diffusion profile of the dye. Middle and bottom rows: Effect of the radius of curvature of the channel. $Q=14$~\textmu l/min, $R_i=135$~\textmu m (middle row) and $R_i=420$~\textmu m (bottom row). Although the relative magnitude of the secondary flow is weaker than in the smaller channel, the larger $Q$ and the integration over a longer distance make the vortical structure clear. The similarity of the interface evolution for both radii supports the scaling of the secondary flow with $De$. }
\label{fig:6}
\end{figure*}

Except at the highest flow rates (or $De^*$), the magnitude of the secondary flow velocities is very small and thus difficult to resolve with PIV techniques. However, if the effect of these small velocities can be integrated over a large distance any effect should be magnified. One method to achieve this integrated effect is through the use of confocal microscopy in combination with dyed stream visualisation, which we now turn our attention to. For those experiments, the fluid supplied through one of the inlets is dyed with fluorescein. The location of the dyed stream is identified in Fig.\ref{fig:2}. At the Y-junction, the two streams each occupy half of the channel width, separated by a straight centred interface in the plane of the channel cross-section. This interface is broadened by diffusion as the fluid travels downstream, and deformed in the region of the loops as the vortices of secondary flows transport fluid in the plane of the cross-section. Following the evolution of this interface by taking slices in the $yz$ plane is thus a means of visualising the fluid transport that has occurred in the cross-sectional plane between consecutive slices. This evolution is shown in Fig.\ref{fig:6} for three channels with different inner radii of curvature, at the six locations identified in Fig.\ref{fig:2}. The interface between the two streams of fluid is quite broad,
due to molecular diffusion of the dye, but also due to the convolution of the image with a finite-sized point spread function, enhanced by the strong illumination conditions. Therefore, only the qualitative evolution of the interface can be obtained, by adjusting the light intensity at each $z$ position with a diffusion-type profile. The inflection point of this profile provides an estimate of the location of the interface, which is marked by the bright lines in Fig.\ref{fig:6}. This diffusion profile is wider towards the top and bottom, suggesting that Taylor dispersion is active in our system~\citep{ismagilov2000}. Coupled with the weaker light intensity close to the walls, this effect is responsible for a loss of resolution at the top and bottom wall, causing the slight bending of the interface observed in the straight channel (location S). 

The top row in Fig.\ref{fig:6} shows the evolution of the interface in the channel we used for the \textmu PIV measurements, at the largest $De^*$ we investigated (0.86). This evolution is in good qualitative agreement with the numerically uncovered nature of the secondary flow as illustrated in Fig.\ref{fig:3}(a): between location S and A0, the fluid has travelled a quarter loop with the dyed stream at the inner edge of the bend. The convex shape of the interface at A0 indicates that the dye has been transported towards the outer edge in the centreplane, and that un-dyed fluid has been carried towards the inner edge at the top and bottom walls. This is consistent with the transport expected from the numerically computed vortex structure. From A0 to A1, transport along a quarter loop of reversed curvature leads to the recovery of a straight interface, which is bent to a concave shape after further transport in the same half loop to location A2. The channel then reverses curvature again, and a convex interface is observed after transport over half a loop (location A4). The images at A0, A2 and A4 are taken at the connection between consecutive half-loops, were the channel reverses curvature. An additional secondary flow is expected to be triggered from this sudden change in curvature even in a Newtonian fluid~\citep{guglielmini2011}. However, the displacement of the interface is not sensitive to the local velocity field, but rather to the integration of the streamlines over the distance travelled in the channel. Therefore, we expect that this is the reason we do not seem to observe the signature of this secondary flow in the cross-sectional profiles measured.

We will now use confocal visualisation to probe the influence of the radius of curvature of the channel $R_i$ on the secondary flow, to gather further experimental proof of the flow scaling with $De$. As the relative magnitude of the secondary flow decreases for larger $R_i$, direct velocity measurements are difficult for those geometries. However, the integration over long distances makes the secondary flow visible in confocal experiments. The second and third rows in Fig.\ref{fig:6} show the evolution of the dyed and un-dyed streams interface in two channels of larger radii of curvature: $R_i=135$~\textmu m (middle row) and $R_i=420$~\textmu m (bottom row), but similar cross-section as the previous channel. $Q_{insta}$ depends non-linearly on $R_i$, therefore for those two larger channels we do not work in terms of quantities scaled on the critical values, such as $De^*$. We keep the flow rate constant, so that the ratio of the Deborah numbers for both experiments is inversely proportional to that of the channel radii: $De_1/De_2 = R_{i,2}/R_{i,1}$. Confocal imaging of the cross-section shows clear evidence of the cross-sectional vortices, with marked deflections of the interface. The interface profiles obtained with the two larger channels have similar curvature, which provides semi-quantitative experimental evidence for the scaling of $u_y$ with $De$: the displacement of the interface between, for instance, S and A0, is proportional to $u_y \times \Delta t$, where $\Delta t$ is the time required for the base flow to travel from S to A0. $\Delta t \sim  R_i/U  \sim  R_i/Q$ because the channels have the same cross-section. Therefore, the lateral displacement of the interface scales as $u_y \times R_i/Q$. As this displacement is similar for the two channels, and $Q$ is identical, $u_y$ scales as $1/R_i$, which is consistent with the linear scaling of $u_y$ with $De$ as measured numerically. 

Finally, we also note that in all cases, the shape of the interface is almost unchanged after transport over an even number of consecutive half-loop (see A0 compared with A4). We thus confirm experimentally for Deborah numbers below one, memory effects are small in our system, as predicted numerically~\citep{Poole2013}.



\section{Conclusions and outlook}
\label{sec:conc}
The use of two complementary techniques, \textmu PIV in the centreplane of the microfluidic device and confocal microscopy to image the cross-section of the device, has allowed us to perform one of the first experimental characterizations of steady, viscoelastic secondary flows in curved microchannels. The vortical structure of this flow in the cross-sectional plane, first unveiled by numerical calculations, was confirmed. Qualitative agreement is found in the flow profiles for the secondary transverse velocity. Those results improve our comprehension of viscoelastic flows in complex channel geometries, by validating the three-dimensional flow driven by the hoop stress in regions of constant curvature. A full understanding of the flow pattern in the serpentine channel, though, remains beyond the scope of our work: in the regions where the curvature is not constant (as is typically the case between consecutive half-loops), additional vortices may appear, which we do not discuss here. Their contribution to the flow dynamics in the serpentine microchannel may be important, though, via their interaction with the viscoelastic Dean flow we have characterised, and their position at a potentially critical location for the propagation of the elastic instability in the channel.

\begin{acknowledgements}

A.L. and L.D. acknowledge funding from the ERC Consolidator Grant PaDyFlow (Grant Agreement no. 682367). R. J. P. acknowledges funding for a ``Fellowship'' in Complex Fluids and Rheology from the Engineering and Physical Sciences Research Council (EPSRC, UK) under grant number EP/M025187/1, and support from Chaire Total. S.J.H., A.Q.S. and L.D. gratefully acknowledge the support of the Okinawa Institute of Science and Technology Graduate University (OIST) with subsidy funding from the Cabinet Office, Government of Japan, and funding from the Japan Society for the Promotion of Science (Grant Nos. 17K06173, 18H01135 and 18K03958). S.L. acknowledges funding from the Institut Universitaire de France. We would also like to acknowledge discussions on the nature of the secondary flow with Philipp Bohr and Christian Wagner. This work has received the support of Institut Pierre-Gilles de Gennes (\'Equipement d'Excellence, ``Investissements d'avenir", program ANR-10-EQPX-34).

\end{acknowledgements}

\bibliographystyle{spbasic}      

\end{document}